\documentclass[prc,twocolumn,showpacs,preprintnumbers,superscriptaddress,
               amsmath,amssymb]{revtex4}
\usepackage{amsmath}           
\usepackage{amsfonts,mathrsfs}
\usepackage{graphicx}          
\usepackage{dcolumn}           

\begin{document}

\preprint{\fbox{\sc version of \today}}

\title{Continuum and Symmetry-Conserving Effects in Drip-line Nuclei Using 
Finite-range Forces}
       
\author{N. Schunck}
\affiliation{\it Departamento de Fisica Teorica, Universidad Autonoma de Madrid
                 28 049 Cantoblanco, Madrid, Spain\\
                   }
\affiliation{\it 
University of Tennessee, Physics Division, 401 Nielsen Physics, Knoxville, 
TN 37996, USA \\
Oak Ridge National Laboratory, Bldg. 6025, MS6373, P.O. 6008, 
Oak Ridge, TN-37831, USA\\
                   }
\author{J. L. Egido}
\affiliation{\it Departamento de Fisica Teorica, Universidad Autonoma de Madrid
                 28 049 Cantoblanco, Madrid, Spain\\
            }
\date{\today}

\pacs{PACS numbers: 21.10.Dr, 21.10.Gv, 21.30.Fe, 21.60.Jz}

\begin{abstract}
We report the first calculations of nuclear properties near the drip-lines 
using the spherical Hartree-Fock-Bogoliubov mean-field theory with a 
finite-range force supplemented by continuum and particle number projection 
effects. Calculations were carried out in a basis made of the eigenstates of a 
Woods-Saxon potential computed in a box, thereby garanteeing that continuum 
effects were properly taken into account. Projection of the self-consistent 
solutions on good particle number was carried out after variation, and an 
approximation of the variation after projection result was used. We give 
the position of the drip-lines and examine neutron densities in neutron-rich
nuclei. We discuss the sensitivity of nuclear observables upon continuum and 
particle-number restoration effects.
\end{abstract} 

\pacs{PACS numbers: 21.10.Dr, 21.10.Gv, 21.10.Re, 21.30.Fe, 21.60.Ev, 21.60.Jz}

\maketitle

%
%

One of the main challenges in contemporary nuclear structure is to describe
nuclei far from the valley of stability. New radioactive ion beam facilities are
under intense development and when in full exploitation will considerably 
increase the wealth of experimental data available \cite{Review-1}. From a 
theoretical point of view, neutron-rich or -deficient nuclei present two major 
challenges. Firstly, these nuclei are weakly-bound, and it is expected that the 
influence of the continuum of energy will significantly alter many nuclear 
properties \cite{Review-2}. However, the extent to which these properties will 
change due to the presence of the continuum is not so clear. Secondly both the 
nuclear mean-field theory and the nuclear shell model rely on effective 
interactions whose parameters were mostly adjusted on data in stable nuclei. 
The reliability and the robustness of these parametrizations under extreme 
isospin ratios must be questioned, and nuclei far from stability thus present 
us with a unique test-ground \cite{Review-1,Review-2}.

Much effort was devoted in the recent years to the study of weakly-bound nuclei, 
and the development of the corresponding appropriate methods, within the 
framework of the non-relativistic mean-field theory with Skyrme interactions 
\cite{Doba-continuum,THO,grasso}, relativistic mean-field \cite{RMF} and the 
nuclear shell model \cite{SMEC}. In mean-field approaches, it was recognized 
early that the stability of very neutron-rich nuclei is caused to a great 
extent by pairing correlations, and that a fully self-consistent treatment of 
the latter in the Hartree-Fock-Bogoliubov formalism is required 
\cite{Doba-continuum}. In the case of the Skyrme mean-field the zero-range of 
the interaction leads to the well-known divergence problem in the pairing 
channel, and practical calculations must resort to phenomenological 
regularization procedures \cite{Pair-reg}. Moreover, approaches based on the 
effective density-functional theory allow terms to be present in the functional, 
that do not originate from a two-body Hamiltonian. The price to pay is the 
occurrence of divergences when going beyond mean-field and restoring the 
particle number \cite{PNP-Gogny,PNP-Diver}. These difficulties do not exist when 
finite-range effective interactions are used \cite{PNP-Gogny}. However, as of 
today, the description of weakly-bound nuclei in the framework of finite-range 
mean-field theory was not available, as existing codes did not take into 
account continuum effects.

In this paper, we present the first attempt to extend the mean-field approach
based on finite-range interactions of the Gogny type to weakly-bound nuclei, 
by simultaneously taking into account continuum effects and the particle number 
symmetry. In section I we describe our method, which relies on the expansion 
of the mean-field solutions on a realistic basis. In section II is shown a 
selection of results benchmarking our method and illustrating the influence of 
the continuum in weakly-bound nuclei calculations. Projection on good particle 
number is discussed in section III.

%
%

\section{Finite-range Mean-Field Theories with Coupling to the Continuum}

Mean-field calculations with finite-range interactions are most conveniently 
carried out in configuration space. In the HFB theory quasi-particle wave-functions 
are thus expanded on a basis of well-known functions. In nuclear structure, the 
harmonic oscillator basis has always played a special role, as it produces 
localized eigenstates that are known analytically. However, a well-known 
deficiency of this basis is its improper asymptotic behavior. While continuum 
states are spherical waves and therefore decay exponentially, HO functions behave 
like Gaussian functions. Unless the number of basis functions is prohibitively 
large, it is therefore impossible to describe all features of weakly-bound nuclei 
with such basis states. 

Apart from the techniques already mentioned \cite{Doba-continuum,THO,grasso,RMF}, 
one simple and elegant way to overcome this problem is to use the eigenstates 
of a "realistic" nuclear potential, that is, of a potential that tends to 0 at 
infinity. Such a potential generates both a sequence of strongly localized 
bound-states and of unlocalized positive-energy states. In practical 
calculations, boundary conditions must be set. Outgoing wave boundary 
conditions with wave number $k$ assure that all solutions with positive energy 
are taken into account (within the indispensable discretization of the $k$ 
values in practical numerical calculations), including resonant and 
non-resonant states. However, the wave-functions are not square-integrable, and 
special techniques must be employed to introduce the completeness relation and 
calculate the matrix elements of the interaction on such a basis 
\cite{nonHermitian-1,nonHermitian-2,nonHermitian-3,nonHermitian-4}. On the 
other hand, vanishing boundary conditions, alternatively named box boundary 
conditions, boil down to selecting only those positive-energy states that have 
a node on the boundaries of the domain of integration. Consequently, not 
the full continuum is taken into account, but the wave-functions are 
orthonormal. For nuclear structure applications and the description of 
ground-state properties the box technique is sufficient and in fact equivalent 
to the full continuum \cite{BoxVsGamow}.

Our model potential was of the Woods-Saxon type, which reads in spherical 
symmetry:
\begin{equation}
V(r) = - \frac{V_{0}}{1 + \exp[(r-r_{0})/a_{0}]}
\label{WS_potential}
\end{equation}
The analytical solutions to the Schr\"odinger equation with this potential are 
not known apart from s-waves states \cite{WSExact}. However, they can be 
determined easily by well-established numerical integration methods. We enclose 
the potential in a sphere of radius $R_{box}$, discretize the space with a mesh 
size $h$ and integrate the radial Schr\"odinger equation numerically: 
\begin{equation}
\frac{d^{2}y_{n\ell}}{dr^{2}} + \frac{2m}{\hbar^{2}}
\left[ e_{n\ell} - \frac{\ell(\ell + 1)}{2mr^{2}} - V(r) \right]y_{n\ell}(r) = 0
\end{equation}
Our tests showed that a mesh size of $h\sim 0.1$ fm guaranteed the convergence 
of the subsequent HFB calculation to within a few keV. 

The parameters of the potential do not play a major role here, as the
eigenfunctions are only used as a basis. Therefore, the particular form of the
Woods-Saxon potential can only affect the speed at which convergence is reached 
but not the final HFB result. We noticed that in order to obtain a fast 
convergence across the whole mass table, the potential should contain a 
consequent amount of bound-states. We therefore adopted the following 
parameters, that correspond to a realistic parametrization of the nuclear 
mean-field of element $Z=126$ \cite{WSUniversal}: $V_{0} = -41.619\ \text{MeV}$, 
$r_{0} = 1.000\ \text{fm}$, $a_{0} = 1.693\ \text{fm}$.

Of more importance is the cut-off of the basis. Like in every numerical
implementation of the mean-field theory, the basis can not be infinite but
must be limited artificially. In practice, our basis functions are characterized
by two quantum numbers, $\ell$ and $n$. Contrary to the harmonic oscillator
basis, where the main shell number $N$ provides a convenient and physical 
cut-off criterion, in the case of the most general potential, the existence of
such a simple criterion can not be guaranteed. Numerical tests showed that in 
heavy nuclei, $\ell_{max} = 15$ and $n_{max} = 18$ allowed very accurate 
calculations from drip-line to drip-line to within 10 keV. For the same 
required accuracy, calculations performed with an energy cut-off (all basis 
states with eigenvalues lower than $E_{cut}$ are retained) reached convergence 
for $E_{cut}$ of the order of 400 MeV. About 92\% of such a basis is made of 
positive-energy unlocalized continuum states.

The Gogny interaction is a finite-range force that contain a central or 
Brink-Boeker term, a spin-orbit, density-dependent and Coulomb terms. The
spin-orbit and Coulomb terms are the same as for zero-range Skyrme forces 
\cite{Gogny}. Only a few parametrizations of this interaction exist. All of 
our calculations were carried out with the D1S force \cite{D1S}. We will study 
in a forthcoming paper the behavior of the isovector part of the Gogny 
interaction in the region of the neutron-drip-line.

\begin{figure}[h]
\hspace{-0.5cm}
\includegraphics[height=6.5cm,width=4.5cm]{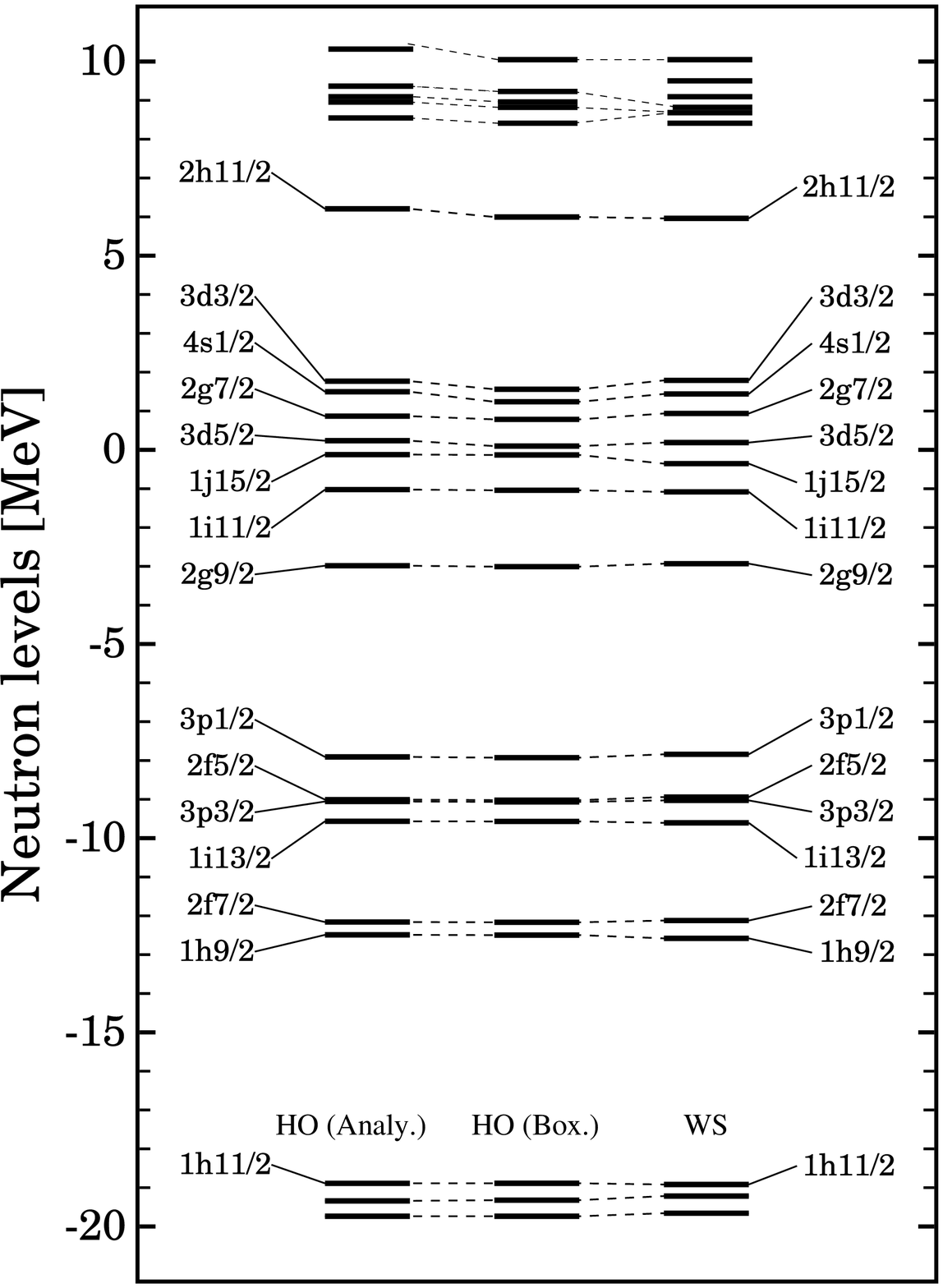}\hspace{-0.4cm}
\includegraphics[height=6.5cm,width=4.5cm]{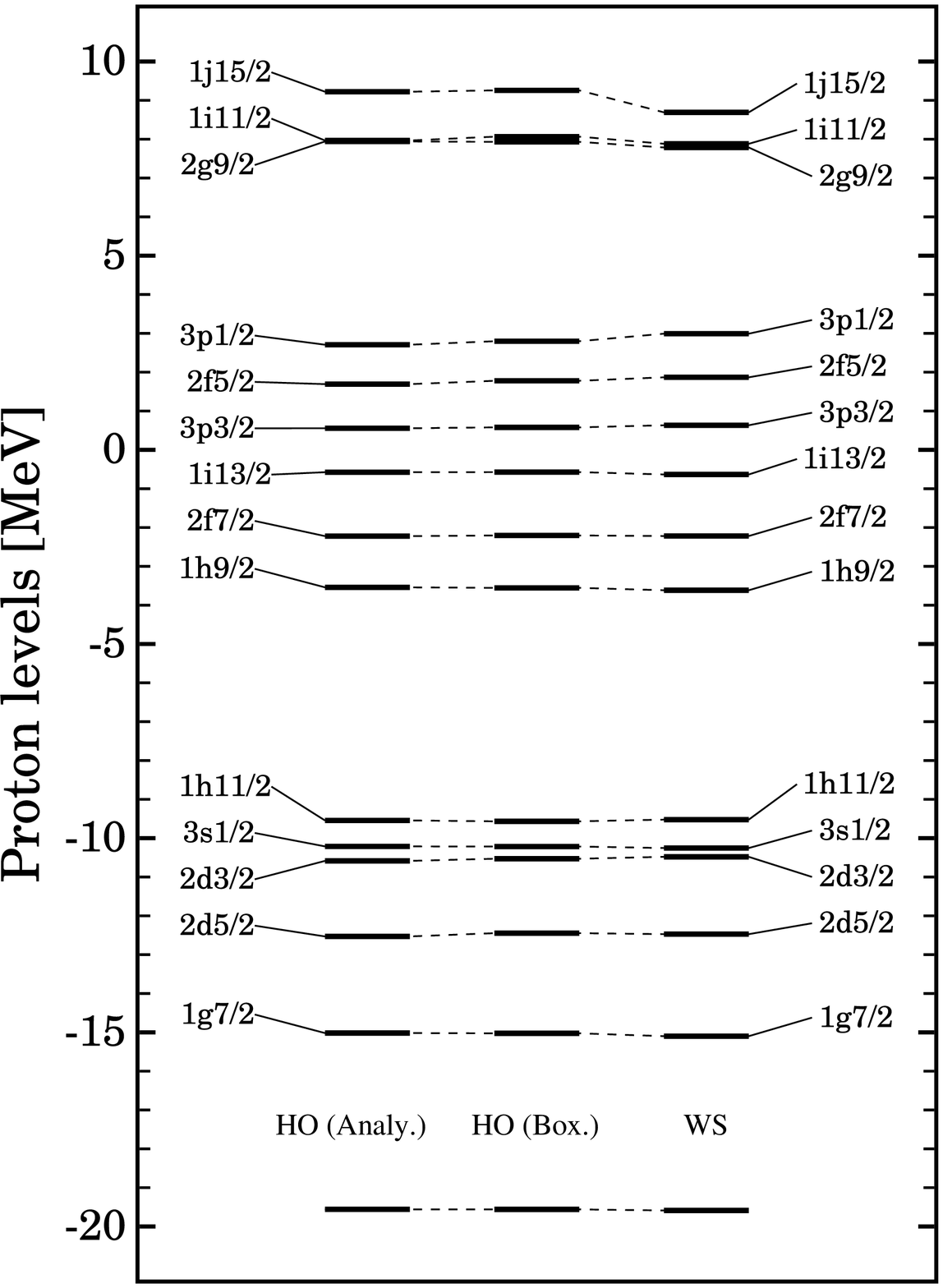}\hspace{-0.6cm}
\caption{
Neutron (left) and proton (right) single-particle energies in 
	$^{208}$Pb calculated in the Hartree-Fock approach with the D1S Gogny 
	effective interaction ($\ell_{max} = 15$ and $n_{max} = 18$). 
	The first two columns in each figure correspond to the harmonic 
	oscillator basis with $N=24$ shells, with either Gauss-Hermite 
	('Analy.') or numerical ('Box.') integrations. The column on the right 
	corresponds to the Woods-Saxon basis.
	 }
\label{fig01}	 
\end{figure}
Figure \ref{fig01} shows the single-particle spectrum in $^{208}$Pb around the
Fermi level as function of the type of basis used for the calculation. 
Benchmark calculations were performed in the HO basis using standard 
Gauss-Hermite quadrature formulas for the integrations (left column) 
\cite{llarena}. The middle column shows the results when the HO basis 
wave-functions are obtained by integrating the Schr\"odinger equation for a HO 
potential in a box and using numerical integration techniques to compute the 
matrix elements of the Gogny interaction. Results obtained in the WS basis are 
shown in the column on the right. As required, all three calculations give 
nearly identical results for bound-states, within the numerical accuracy of 
the integration. 

Let us notice that, as the convergence in the HO basis depends on the number of 
shells and of the oscillator length, the convergence in the WS basis depends on 
the cut-offs $\ell_{max}$ and $n_{max}$ (or alternatively $E_{cut}$), on the 
parameters of the WS potential itself {\em and} on the size of the box. A
detailed study of the convergence properties of the WS basis will be presented
elsewhere.

%
%

\begin{table}[h]
\begin{center}
\caption{Table of Fermi level and separation energies in drip-line nuclei}
\begin{ruledtabular}
\begin{tabular}{ccc|ccc}
\multicolumn{3}{c}{Neutrons} & \multicolumn{3}{c}{Protons}\\
 & $\lambda_{n}$ & $-S_{2n}$ &  & $\lambda_{p}$ & $-S_{2p}$ \\
\hline
$^{~20}$C  & -1.39 & 2.92 & $^{~10}$C  & -3.78 & 11.78\\
$^{~26}$O  & -0.36 & 1.16 & $^{~14}$O  & -4.84 & 11.96 \\
$^{~30}$Ne & -0.88 & 3.78 & $^{~18}$Ne & -2.38 & 5.18 \\
$^{~40}$Mg & -0.47 & 1.44 & $^{~20}$Mg & -1.09 & 2.66 \\
$^{~46}$Si & -0.43 & 1.50 & $^{~24}$Si & -2.06 & 6.24 \\
$^{~50}$S  & -0.17 & 1.30 & $^{~28}$S  & -1.20 & 5.22 \\
$^{~56}$Ar & -0.59 & 1.46 & $^{~32}$Ar & -1.46 & 4.18 \\
$^{~64}$Ca & -0.09 & 0.06 & $^{~36}$Ca & -1.01 & 5.56 \\
$^{~72}$Ti & -0.18 & 0.66 & $^{~40}$Ti & -0.27 & 1.78 \\
$^{~76}$Cr & -0.16 & 0.62 & $^{~44}$Cr & -1.21 & 3.24 \\
$^{~82}$Fe & -0.26 & 0.86 & $^{~46}$Fe & -0.17 & 1.02 \\
$^{~86}$Ni & -0.22 & 0.98 & $^{~52}$Ni & -1.18 & 5.41 \\
$^{~92}$Zn & -0.18 & 0.60 & $^{~58}$Zn & -0.70 & 2.53 \\
$^{104}$Ge & -0.12 & 0.14 & $^{~62}$Ge & -0.77 & 3.22 \\
$^{114}$Se & -0.33 & 0.76 & $^{~66}$Se & -0.89 & 3.54 \\
$^{116}$Kr & -1.10 & 2.24 & $^{~68}$Kr & -0.08 & 0.71 \\
$^{120}$Sr & -0.44 & 2.76 & $^{~72}$Sr & -0.54 & 1.88 \\
$^{122}$Zr & -1.02 & 4.36 & $^{~78}$Zr & -1.12 & 3.37 \\
$^{130}$Mo & -0.02 & 0.16 & $^{~82}$Mo & -0.38 & 2.90 \\
$^{136}$Ru & -0.01 & 0.14 & $^{~86}$Ru & -0.63 & 2.12 \\
$^{140}$Pd & -0.08 & 0.42 & $^{~90}$Pd & -0.86 & 2.45 \\
$^{152}$Cd & -0.03 & 0.04 & $^{~94}$Cd & -1.05 & 2.66 \\
$^{170}$Sn & -0.03 & 0.12 & $^{102}$Sn & -0.70 & 6.06 \\
$^{176}$Te & -0.44 & 0.90 & $^{110}$Te & -0.04 & 1.53 \\
$^{178}$Xe & -0.99 & 1.98 & $^{114}$Xe & -0.19 & 1.43 \\
$^{180}$Ba & -1.51 & 3.02 & $^{118}$Ba & -0.37 & 1.62 \\
$^{184}$Ce & -0.26 & 3.16 & $^{122}$Ce & -0.50 & 1.80 \\
$^{186}$Nd & -0.65 & 4.10 & $^{126}$Nd & -0.52 & 1.79 \\
$^{188}$Sm & -1.08 & 5.06 & $^{130}$Sm & -0.41 & 1.59 \\
$^{190}$Gd & -1.52 & 6.00 & $^{134}$Gd & -0.04 & 1.32 \\
$^{198}$Dy & -0.10 & 0.38 & $^{140}$Dy & -0.34 & 1.39 \\
$^{206}$Er & -0.02 & 0.04 & $^{144}$Er & -0.19 & 1.14 \\
$^{220}$Yb & -0.02 & 0.02 & $^{148}$Yb & -0.11 & 0.90 \\
$^{240}$Hf & -0.03 & 0.12 & $^{152}$Hf & -0.05 & 0.72 \\
$^{252}$W  & -0.06 & 0.20 & $^{158}$W  & -0.59 & 1.76 \\
$^{258}$Os & -0.15 & 0.32 & $^{162}$Os & -0.49 & 1.48 \\
$^{260}$Pt & -0.44 & 0.88 & $^{166}$Pt & -0.38 & 1.17 \\
$^{262}$Hg & -0.71 & 1.42 & $^{170}$Hg & -0.25 & 0.79 \\
$^{264}$Pb & -1.00 & 1.46 & $^{182}$Pb & -0.08 & 4.12 \\
$^{268}$Po & -0.00 & 1.70 & $^{194}$Po & -0.16 & 1.84 \\
$^{270}$Rn & -0.24 & 2.22 & $^{198}$Rn & -0.07 & 1.11 \\
$^{272}$Ra & -0.50 & 2.72 & $^{204}$Ra & -0.43 & 1.70 \\
$^{274}$Th & -0.76 & 3.22 & $^{208}$Th & -0.21 & 1.26 \\
$^{276}$U  & -1.04 & 3.76 & $^{214}$U  & -0.32 & 1.56 \\
$^{282}$Pu & -0.06 & 0.28 & $^{220}$Pu & -0.42 & 1.54 \\
$^{296}$Cm & -0.04 & 0.04 & $^{222}$Cm & -0.17 & 1.11 \\
\end{tabular}
\label{table01}
\end{ruledtabular}
\end{center} 
\end{table}

\section{Drip-lines and Continuum Effects}

As an application of the consideration of continuum effects in practical 
calculations we present in Table \ref{table01} the neutron and proton 
drip-lines calculated in the spherical HFB approach for the Woods-Saxon basis. 
For the listed elements, the two-neutrons $S_{2n}$ and two-protons $S_{2p}$ 
separation energies were calculated, as well as the position of the Fermi level. 
The criterion to define the drip-line was either a positive separation energy, 
or a positive Fermi level, whichever criterion was met first. The WS basis 
included all states with $\ell\leq 15$ and $n\leq 18$ in a box of $R_{box}=20$ 
fm. 

Drip-lines can be significantly pushed away by long-range correlations. In 
particular, rare-earth nuclei are found to be strongly deformed either in 
large-scale Gogny HFB calculations using the D1S interaction 
\cite{Gogny-MassTable} and including beyond mean-field correlations 
\cite{RE-07} as well as in Skyrme HFB calculations performed in the transformed 
harmonic oscillator basis with the SLy4 parametrization in the mean-field
channel and the volume delta pairing force \cite{THO-mass}. Bearing this fact 
in mind, we nevertheless observe systematic differences on the position of the 
neutron drip-line as compared to Skyrme HFB calculations published in 
\cite{THO-mass} where the drip-line almost systematically extends further away 
by 2-4 neutrons.

\begin{figure}[h]
\includegraphics[height=7.0cm,width=9.0cm]{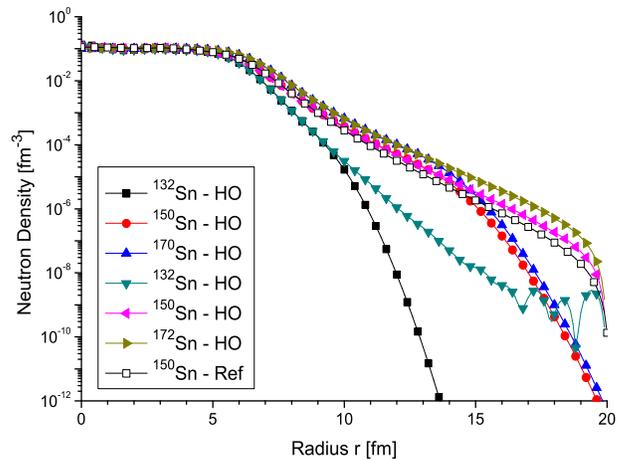}
\caption{(color online) Neutron densities in neutron-rich $^{132,150,170}$Sn 
isotopes computed with the finite range D1S Gogny interaction in the HO and 
Woods-Saxon bases for a box of 20 fm. The curve marked Ref. is the HFB 
calculation in coordinate space for $^{150}$Sn with the SkP force 
\cite{Doba-continuum}. The unphysical asymptotic behavior of densities 
computed in the HO basis is clearly visible. Oscillations in the case of 
$^{132}$Sn are caused by the absence of pairing correlations which forbids 
$\ell$-couplings in the expansion.
	 }
\label{fig02}	
\end{figure}
Observables like densities are very sensitive to continuum effects and 
significantly impact many spectroscopic or reaction properties of nuclei. A 
stringent test of our method is therefore its ability to reproduce the long
tails of neutron densities in neutron-rich nuclei. There have been numerous
studies of neutron skins in exotic nuclei in the past, and we can therefore
benchmark our approach.

As an example we display in figure \ref{fig02}, the neutron densities of three 
very neutron-rich Sn isotopes. For small and medium values of the nuclear 
radius the HO and the WS bases give the same density profile. However, for 
large radius the differences are significant. As expected, the proper 
asymptotic of the WS basis wave-functions is reflected in the long density 
tails. It is important to emphasize that the results are nearly identical to 
coordinate space Skyrme HFB calculations of \cite{Doba-continuum} as can be 
seen from the curve marked Ref., which shows the density in $^{150}$Sn
obtained with the SkP parametrization of the Skyrme interaction. This provides 
an independent validation and very robust test of our method to incorporate
continuum effects through a basis.

We observe oscillations in the densities for closed-shells nuclei like
$^{132}$Sn. Indeed, in that case the density is simply the sum of $N$ 
well-localized single-particle states, and no contribution from discretized 
continuum states enters the expansion. The threshold where these oscillations 
appear gives in fact a measure of the intrinsic numerical accuracy with which
integrations are performed. This phenomenon may be distinguished from what was 
reported in \cite{OscillHFB} where oscillations were caused by large 
discrepancies between the upper and lower components in the HFB equations.

%
%

\section{Restoration of the Particle Number Symmetry}

As mentioned in the introduction pairing correlations play a relevant role in 
the description of weakly bound nuclei. Up to now most descriptions of the 
drip-line were performed in the HFB approach which has two drawbacks. Firstly, 
in the weak pairing regime it usually collapses to the unpaired solution and 
secondly, the HFB wave function has the right number of particles only on the 
average. To restore this spontaneously broken symmetry, we project the HFB 
solution onto good particle number using the projector:
\begin{equation}
\hat{P}^{N} = \frac{1}{2\pi}\int_{0}^{2\pi} d\varphi e^{i\varphi(\hat{N} - N)}  
\end{equation}
according to the procedure described in details in \cite{PNP-Gogny}. As is 
well-known, the variation after projection (VAP) method provides a much better 
approach than the projection after variation (PAV) at the cost of a much 
higher numerical effort, especially with the huge basis used in the present 
calculations. In particular, it prevents pairing correlations to collapse. 
Nevertheless one can obtain an approximate VAP solution by 
searching for the minimum in a restricted highly pair-correlated variational 
space \cite{restricted-VAP}. This space is generated by HFB solutions 
$|\phi(x)\rangle$ corresponding to the self-consistent diagonalization of the 
one-body Hamiltonians $\hat{h}_{HFB}(x)$, as a function of $x$. Here 
$\hat{h}_{HFB}(x)$ stands for the HFB Hamiltonian obtained by multiplying, at
each iteration, the pairing field $\Delta$ of the HFB matrix by the constant $x$, i.e., 
$\Delta \rightarrow \Delta(x) = x\Delta$, while the particle-hole field $h$ 
remains unchanged. To each wave function $|\phi(x)\rangle$  
is associated an eigenstate of the particle number operator 
$|\Phi(x)\rangle = \hat{P}^{N}|\phi(x)\rangle $ and an energy 
\begin{equation}
E^{N}(x) =
\frac{\langle\Phi(x)|\hat{H}|\Phi(x)\rangle}
{\langle\Phi(x)|\Phi(x)\rangle}
\end{equation}
with $\hat{H}$ the {\em original} two-body Hamiltonian. The variational principle 
guarantees that such  a procedure yields a minimum as function of $x$, which is 
an excellent approximation to the VAP result \cite{restricted-VAP}. We call 
this approach RVAP (Restricted VAP).
\begin{figure}[h]
\includegraphics[height=7.00cm,width=9.0cm]{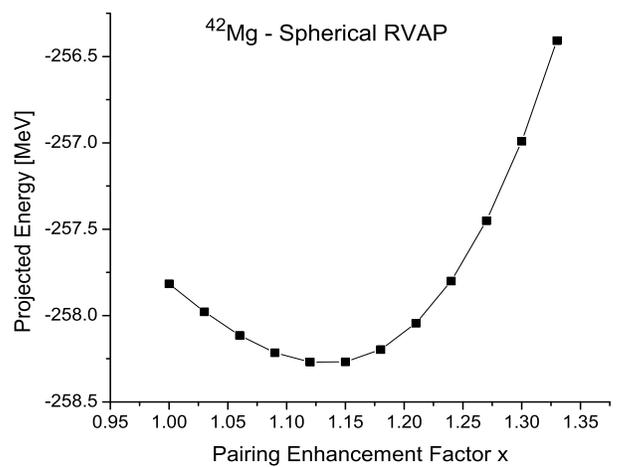}
\caption{(color online) Total projected energy $E^{N}(x)$ in $^{42}$Mg as function 
of the effective pairing strength factor $x$. The RVAP calculation was performed in 
the WS basis.}
\label{fig03}	 
\end{figure}

To investigate the impact of a better treatment of pairing correlations on
exotic nuclei we applied this restricted VAP procedure in the neighborhood of 
the neutron drip-line. Figure \ref{fig03} shows the example of $^{42}$Mg which 
is unbound against 2 neutrons emission in standard spherical HFB or PAV-HFB calculations. 
The minimum of the curve depicted in Fig. \ref{fig03} corresponds to a 
stronger pairing regime than in standard HFB calculations: the total gain in 
energy is only 460 keV but the pairing energy goes from -9.22 MeV to -13.83 MeV 
from $x=1.0$ to $x=1.12$. At the minimum, the two-neutron separation energy 
calculated from the RVAP projected energies of $^{42}$Mg and $^{40}$Mg becomes 
negative at $S_{2n} = -252$ keV (compared with $S_{2n} = + 40$ keV in the 
un-projected case) and the nucleus is bound against 2 neutrons 
emission. This effect is usually typical from nuclei near closed-shells 
where only the VAP method can prevent the HFB theory to collapse to the 
unpaired regime. Near the drip-lines, it also results in a visible shift of the 
drip-lines. It is important to realize that this effect can not appear in a 
PAV calculation.

The restoration of symmetries broken at the mean-field level has therefore a
significant and experimentally visible effect in exotic neutron-rich nuclei. 
Variation after projection on good particle number consists in exploring a 
different variational space than when the projection is carried out after the 
variation.

In summary, we present the first results of a generalization of HFB 
calculations with finite-range interactions to exotic weakly-bound nuclei. 
Mean-field solutions are expanded on a basis of the eigenstates of a 
Woods-Saxon potential, which possess the proper asymptotic behavior. In stable 
nuclei, this WS basis is benchmarked against standard harmonic oscillator
calculations and is found to give exactly the same results. In neutron-rich 
nuclei we find that the asymptotic of the neutron densities is properly
reproduced. The role of the restoration of particle number {\em before 
variation} is emphasized. It is found that this mechanism can stabilize 
nuclei at the drip-line thereby pushing away the latter.

{\bf Acknowledgment - } We wish to thank P. Ring and T. R. Rodriguez for 
fruitful discussions. One of us (N.S.) acknowledges financial support of the 
spanish Ministerio de Educacion y Ciencia (Ref. SB2004-0024). This work has 
been supported in part by the DGI, Ministerio de Ciencia y Tecnologia, Spain, 
under Project FIS2004-06697, as well as by the U.S. Department of Energy under 
Contract Nos. DE-FC02-07ER41457 (University of Washington), DEFG02- 96ER40963 
(University of Tennessee), and DE-AC05- 00OR22725 with UT-Battelle, LLC (Oak 
Ridge National Laboratory). We thank K. Bennaceur and J. Dobaczewski for
letting us use the code HFBRAD.

%
%


\begin{thebibliography}{30}


\bibitem{Review-1}
	{
	J. Dobaczewski and W. Nazarewicz
	Phil. Trans. R. Soc. Lond. A {\bf 356}, 2007-2031 (1998)
	}
\bibitem{Review-2}
	{
	J. Dobaczewski, N. Michel, W. Nazarewicz, M. P{\l}oszajczak and J. Rotureau
	Prog. Part. Nucl. Phys. {\bf 56}, 432-455 (2007)
	} 
\bibitem{Doba-continuum}
	{
	J. Dobaczewski, H. Flocard and J. Treiner,
	Nucl. Phys. {\bf A422}, 103 (1984);
	J. Dobaczewski, W. Nazarewicz, T. R. Werner, J.-F. Berger, C. R. Chinn
	and J. Decharg\'{e}, Phys. Rev. {\bf C53}, 2809-2840 (1996)
	}
\bibitem{THO}
	{
	M. V. Stoitsov, W. Nazarewicz and S. Pittel, 
	Phys. Rev. {\bf C58}, 2092 (1998);
	M. V. Stoitsov, J. Dobaczewski, W. Nazarewicz and P. Ring, 
	Comput. Phys. Commun. {\bf 167}, 43-63 (2005)
	}
\bibitem{grasso}
	{
	M. Grasso, N. Sandulescu, N. Van Giai and R. J. Liotta, 
	Phys. Rev. {\bf C64}, 064321 (2001)
	}
\bibitem{RMF}
	{
	W. Poeschl, D. Vretenar and P. Ring,
	Comp. Phys. Comm. {\bf 103}, 217 (1997)
	}
\bibitem{SMEC}
	{
	N. Michel, W. Nazarewicz, M. P{\l}oszajczak and J. Oko{\l}owicz
	Phys. Rev. {\bf C67}, 054311 (2003);
	N. Michel, W. Nazarewicz, M. P{\l}oszajczak and K. Bennaceur
	Phys. Rev. Lett. {\bf 89}, 042502 (2002)
	}
\bibitem{Pair-reg}
	{
	P. Borycki, J. Dobaczewski, W. Nazarewicz and M. Stoitsov
	Phys. Rev. {\bf C73}, 044319 (2006)
	}
\bibitem{PNP-Gogny}
	{
	M. Anguiano, J.L. Egido and L.M. Robledo
	Nucl.Phys. {\bf A696}, 467-493 (2001);
	M. Anguiamo, J.L. Egido and L.M. Robledo
	Phys. Lett. {\bf B545}, 62-72 (2002);
	}
\bibitem{PNP-Diver}
	{
	J. Dobaczewski, M.V. Stoitsov, W. Nazarewicz, P.-G. Reinhard, 
	nucl-th/arXiv:0708.0441 
	}
\bibitem{nonHermitian-1}
	{
	Y.B. Zel'dovich, Sov. Phys. JETP. {\bf 12}, 542 (1960);
	N. Hokkyo, Prog. Theor. Phys. {\bf 33}, 1116 (1965);
	W. J. Romo, Nucl. Phys. {\bf A116}, 617 (1968);
	}
\bibitem{nonHermitian-2}
	{
	T. Berggren, Nucl. Phys. {\bf A109}, 265 (1968);
	}
\bibitem{nonHermitian-3}
	{
	B. Gyarmati and T. Vertse, Nucl. Phys. {\bf A160}, 523 (1971);
	G. Garcia-Calderon and R. Peierls, Nucl. Phys. {\bf A265}, 443 (1976);
	}
\bibitem{nonHermitian-4}
	{
	P. Lind, Phys. Rev. {\bf C47} 1903 (1993)
	}
\bibitem{BoxVsGamow}
	{
	M. Stoitsov, N. Michel and K. Matsuyanagi,
	nucl-th/arXiv:0709.1006
	}
\bibitem{WSExact}
	{
	S. Fl\"{u}gge, Practical Quantum Mechanics I (Springer-Verlag, Berlin, 1971) vol. 1, Prob.
	No. 37, 39.
	}
\bibitem{WSUniversal}
	{
	J. Dudek and T. R. Werner, 
	Phys. Lett. {\bf 10}, 1543 (1978)
	}
\bibitem{Gogny}
	{
	J. Decharg\'e and D. Gogny,
	Phys. Rev. {\bf C21}, 1568 (1980)
	}
\bibitem{D1S}
	{
	J.-F. Berger, M. Girod and D. Gogny, Comput. Phys. Commun. {\bf 63}, 365 (1991)
	}
\bibitem{llarena}
	{
	T. Gonzalez-Llarena, Ph.D Thesis, Universidad Autonoma de Madrid, Sep.
	1999, unpublished.
	}
\bibitem{Gogny-MassTable}
	{
	http://www-phynu.cea.fr/science{\_}en{\_}ligne/carte{\_}poten-\\
tiels{\_}microscopiques/carte{\_}potentiel{\_}nucleaire.htm
	}
\bibitem{RE-07}
	{
	T. R. Rodriguez and J. L. Egido, 
	Phys. Rev. Lett. {\bf 99}, 062501 (2007)
	}	
\bibitem{THO-mass}
	{
	M. V. Stoitsov, J. Dobaczewski, W. Nazarewicz, S. Pittel and D. J. Dean,
	Phys. Rev. {\bf C68}, 054312 (2003)
	}
\bibitem{OscillHFB}
	{
	K. Bennaceur and J. Dobaczewski, 
	Comp. Phys. Comm. {\bf 168}, 96-122 (2005)
	}
\bibitem{restricted-VAP}
	{
	T. R. Rodriguez, J. L. Egido and L. M. Robledo, 
	Phys. Rev. {\bf C72}, 064303 (2005)
	}

\end{thebibliography}
\end{document}